# Machine Learning-based Approach for Depression Detection in Twitter Using Content and Activity Features

Hatoon AlSagri [1,2]
[1]Information Systems Department
College of Computer and Information Sciences
Al Imam Mohammad Ibn Saud Islamic University
Riyadh, Saudi Arabia

Mourad Ykhlef [2]
[2]Information Systems Department
College of Computer and Information Sciences
King Saud University
Riyadh, Saudi Arabia

**Abstract**
Social media channels, such as Facebook, Twitter, and Instagram, have altered our world forever. People are now increasingly connected than ever and reveal a sort of digital persona. Although social media certainly has several remarkable features, the demerits are undeniable as well. Recent studies have indicated a correlation between high usage of social media sites and increased depression. The present study aims to exploit machine learning techniques for detecting a probable depressed Twitter user based on both, his/her network behavior and tweets. For this purpose, we trained and tested classifiers to distinguish whether a user is depressed or not using features extracted from his/her activities in the network and tweets. The results showed that the more features are used, the higher are the accuracy and F-measure scores in detecting depressed users. This method is a data-driven, predictive approach for early detection of depression or other mental illnesses. This study's main contribution is the exploration part of the features and its impact on detecting the depression level.

**Introduction**
Depression is a common mental illness and a leading cause of disability worldwide, which may cause suicides. Globally, more than 300 million people are estimated to suffer from depression every year[1]. Generally, depression is diagnosed through face-to-face clinical depression criteria. However, at early stages of depression, 70% of the patients would not consult doctors, which may take their condition to advance stages [1].
Recently, there has been a movement to leverage social medial data for detecting, estimating, and tracking the changes in occurrence of a disease [2]. The ubiquity of social media provides a rich opportunity to enhance the data available to mental health clinicians and researchers, enabling a better-informed and -equipped mental health field [3]. In addition, contagious negative emotions in social networks adversely affect people, leading to depression and other mental illnesses. Mental illness is known as a major risk factor of suicide; almost 80% of those who attempt or die by suicide are known to have had some form of mental illness [4, 5]. Depression is considered as the most common mental illness [6], but because of its unrecognition or denial, it has remained undiagnosed or untreated [7]. The onset of a major depression can be prevented by early recognition of its symptoms and their treatment through timely intervention [7].
Many studies have detected physical and mental illnesses derived from social media's huge information, in precise some studies were dedicated for depression [3, 8, 9]. De Choudhury et al. [8] found that tweets posted by individuals with major depressive disorder, as well as their social media activity, can be utilized to classify and predict if they are suffering from depression or are likely to suffer in the future. Nadeem et al., Dredze et al., and Benton et al. focused on whether the users' tweets were depressive in nature [10-12], while Tsugawa et al. and Coopersmith et al. focused on user activity in Twitter [3, 13].

---

1 (https://www.who.int/en/news-room/fact-sheets/detail/depression)

This study also aims to detect whether the user is depressed, from the nature of his/her tweets and activity in the network. It can be further used to identify other mental illnesses and might even form an underlying infrastructure for new mechanisms that would help detect and limit depression diffusion in social networks.

This study exploits data collected from 111 user profiles and more than 300,000 tweets. Many classifier techniques are employed to identify the depression level, of which support vector machine (SVM)-linear shows the best results, with the accuracy reaching 82.5 and F-measure reaching 0.79.

The rest of this paper is organized as follows: Section 1 illustrates the literature review. Section 2 presents the scientific background of the classifiers used in the experiments. Section 3 explains the methodology of the study and the features extracted and computed. Section 4 describes the experiments and discusses results. Finally, section 5 outlines the conclusions of the study.

**1. Literature Review**

With the gradual increase in social media usage and the extensive level of self-disclosure within such platforms, efforts to detect depression from Twitter data have increased [9, 14]. Park et al. [15] indicated that depressed Twitter users tend to post tweets containing negative emotional sentiments more than healthy users. In addition, De Choudhury et al. [8] found that depressive signals are noticeable in tweets posted by users with major depressive disorder [9, 10, 16].

Thus far, different features have been used to detect depression from Twitter data. De Choudhury et al. [8] collected more than 2 M tweets from 476 users who were clinically diagnosed as depressed and had Twitter accounts. They used behavioral attributes related to social engagement, emotion, language and linguistic styles, ego network, and mentions of antidepressant medications to build a classifier that provides estimates of the risk of depression. They leveraged these distinguishing attributes to build an SVM classifier that can predict the risk of depression with 70% classification accuracy. Tsugawa et al. [13] revealed that frequencies of word usage, along with topic modeling, are useful features for the prediction model. Using the radial kernel SVM classifier, they obtained 69% classification accuracy in predicting depression of 81 participants out of the 209 collected using a questionnaire. In addition, Reece et al. [9] extracted predictive features for measuring the effect, linguistic style, and context from users' tweets; built models using these features with supervised learning algorithms, and successfully discriminated between depressed and healthy contents. Their data were collected from 105 out of the 204 depressed users and the CESD scores relied on the identification of depressed users. The best classifier performance was obtained using a 1200-tree random forest classifier, increasing the precision to 0.866, compared to other study results. Nadeem et al. [10] utilized the bag-of-words approach for better depression detection, which uses word occurrence frequencies to quantify the content of a tweet measured on a document level. They employed four types of binary classifiers: linear SVM classifier, decision tree (DT), Naïve Bayes (NB) algorithm, and logistic regressive approach, and found that NB outperformed other classifiers with an accuracy of 81% and precision of 0.86. They used a corpus of more than 2.5 M tweets gathered from the Shared Task organizers of CLPsych 2015, online, from users who indicated they were diagnosed as depressed (326) or with PTSD. On the other hand, Nadeem et al. [10], Coppersmith et al. [3], and Mowery et al. [2] considered sentiment analysis as a feature to detect depression from Twitter data. Jamil et al. [14] concluded that the use of sentiment analysis, along with percentage of depressed tweets, increases the precision and recall of detecting depression. The classifier was trained on 95 users who disclosed their own depression (which was equal to 5% of users participating in the study, while the remaining 95% were healthy users), using SVM, which provided a recall of 0.875 and precision of 0.775.

De Choudhury et al. [8] and Jamil et al. [14] used the benefits of depressed people tweets for extracting features that helped increase the detection accuracy. De Choudhury et al. [8] built a depression lexicon of terms that are likely to appear in postings from individuals discussing depression or its symptoms in online settings. In contrast, Jamil et al. [17] used the percentage of depressed tweets, along with self-indication of depression, to decide whether a user should be removed from the training set and found this feature to increase the model's accuracy.

## 2. Background
### 2.1 Classifiers

SVM, Naive Bayes (NB), and Decision Tree (DT are some of the widely used algorithms in natural language processing tasks [17]. Of these, SVM-linear classifier demonstrates the best performance. As there is no one algorithm suited for all tasks, researchers tend to try various algorithms and enhance them for the problem of their interest [17].

*NB:*

NB is based on the "Bayes' theorem" in probability. As a requirement of this theorem, NB can be applied only if the features are independent of each other [18].

$$P(X|Y) = \frac{P(Y|X)P(X)}{P(Y)}$$

It is a prediction model that breaks the posterior possibilities of each class and the possible circumstances of the class for each feature. It is generally used in machine learning owing to its ability to efficiently merge the evidence from several features [19]. Often, we know how frequently some evidence is observed, given a known outcome [17]. With the knowledge that certain evidence is observed provides us a conclusion. Although the NB classifier is considered the most straightforward method in the machine learning field, it is still competitive with SVM [10].

*DT:*

DTs classify instances by sorting them based on the feature values. Each node in a DT represents a feature, and each division represents a value that the node can undertake [20] (Figure (1)).

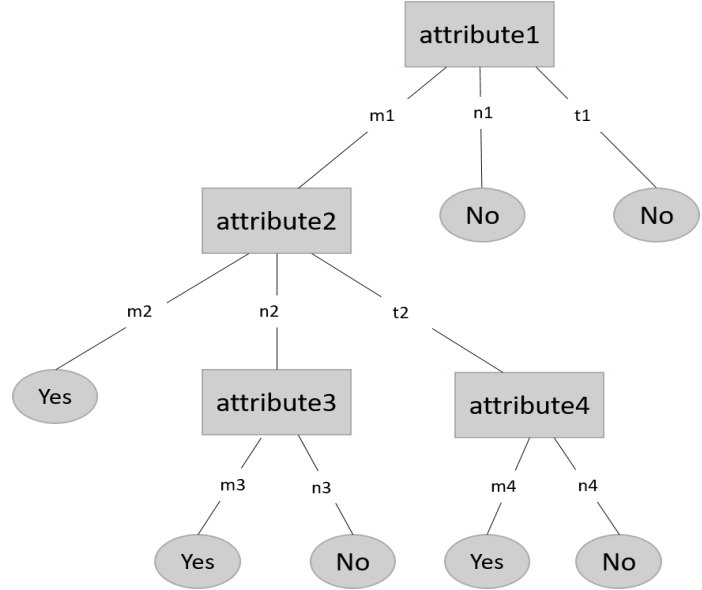
*Figure 1: DT example*

A DT involves partitioning the data into subdivisions that contain occurrences with similar values, using an important aspect in a DT, called split selection, which intends to find an attribute and its related splitting function for each test node in a DT. Splits are evaluated by calculating entropy [21].
Consequently, the entropy E of an attribute *X* of a training object n with the possible attribute values $x_1,...,x_m$ and probability distribution $p(X(n) = x_i)$ is defined as [21]

$$E_X = -\sum_t p_t \cdot \log_2 p_t$$

Let $E_Y$ be the entropy of the decision class distribution, $E_X$ be the entropy of values of an attribute X, and $E_{XY}$ be the entropy of the combined decision class distribution and attribute values:

$$E_Y = -\sum_k p_k \cdot \log_2 p_k$$
$$E_X = -\sum_t p_t \cdot \log_2 p_t$$
$$E_{XY} = -\sum_t \sum_k p_{tk} \cdot \log_2 p_{tk}$$

A DT can structure complicated nonlinear decision borders [22]. The extensive trees will be structured, and then excluded to reduce the cost-complexity criterion. The resulting tree would be easily explicable and can provide perception into the data structure that is claimed to be a main advantage of tree algorithms [17]. DTs simply pose a series of carefully constructed questions to classify a task that makes them straightforward in nature, for which they are extensively employed within the machine learning field [10].

***SVM:***
SVM is a supervised learning model that underlines two different classes in a high-dimensional space. It can adjust several features while balancing the excellent performance, to reduce the possibility of overfitting [23]. SVM is famous for its powerful capability, specifically when working on real-world data, which include a decisive theoretical basis and its insensitivity to high-dimensional data [24, 25]. SVM is a type of algorithm with a set of labeled training examples for a binary class problem. The training algorithm in SVM creates a potential hyperplane, which divides the cases from the two classes. It escalates the distance between the divided hyperplane and the training examples closest to the hyperplane [26]. SVM can provide the prediction and determine which side of the hyperplane an object inclines [17], as shown in Figure (2).

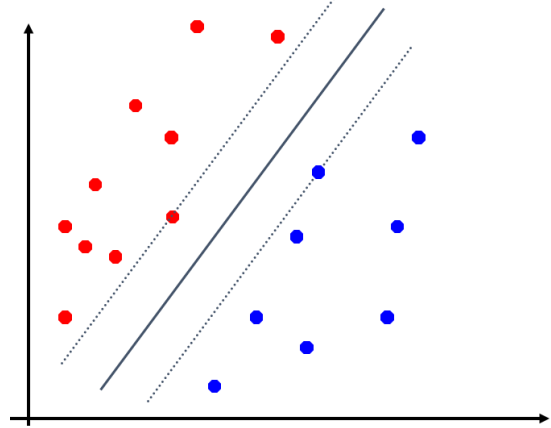

*Figure 2: SVM's maximum separating hyperplane*

The linear classifier relies on the inner product between vectors [20] (support vector $a_i$ and test tuple $a^T$):

$$K(a_i, a_j) = a_i^T a_j$$

If each data point is mapped into high-dimensional space via some transformation Φ: a → φ(a), the inner product becomes

$$K(a_i, a_j) = \varphi(a_i)^T \varphi(a_j)$$

A kernel function corresponds to an inner product in some expanded feature space. A common kernel function is the radial basis function (infinite-dimensional space):

$$K(a_i, a_j) = exp(-\frac{\|a_i - a_j\|^2}{2\sigma^2})$$

Although SVM can tolerate the data outlier, itis computationally inefficient and sensitive to the kernel hyperparameters.

## 3 Methodology

A quantitative study is conducted to train and test various machine learning classifiers to determine whether a twitter account user is depressed, from tweets initiated by the user or his/her activities on Twitter.

Data preparation, feature extraction, and classification tasks are performed using various R packages and in R version 3.3 [27] using Rstudio IDE [28]. The classifiers are trained using 10-fold cross validation to avoid overfitting, and then tested on a held-out test set.

Figure (3) illustrates the depression detection using activity and content features (DDACF) classification model. First, all tweets for depressed and non-depressed accounts, as well as information of user account and activities such as number of followers, number of following, total number of posts, time of posts, number of mentions, and number of retweets, are retrieved. Next, all tweets of an account are assembled in one document.

Text preprocessing is applied to all documents. First, a corpus is created and tweets in each document are tokenized. Next, normalization is applied, where all characters are turned to lower case and punctuations, retweets, mentions, links, unrecognized emoji's, and symbols are removed. Usually, normalization includes removing stop words, such as first-person pronouns like "I," "me," and "you," but when removing stop words, we keep the first-person pronouns. Later, stemming is applied and a document term matrix (DTM) is created for each account [29]. The matrix indicates the frequency of words in each tweet, where each row indicates a document of tweets and each column indicates all words used in all accounts. TF-IDF is used to measure the words' weight.

Features applied on the DTM are then merged with account measures extracted from the social network and user activities. Results of the merge are then treated as independent variables in a classification algorithm to predict the dependent variable of an outcome of interest. Ultimately, we decide upon the DT, a linear and radial kernel support vector classifier, and an NB algorithm.

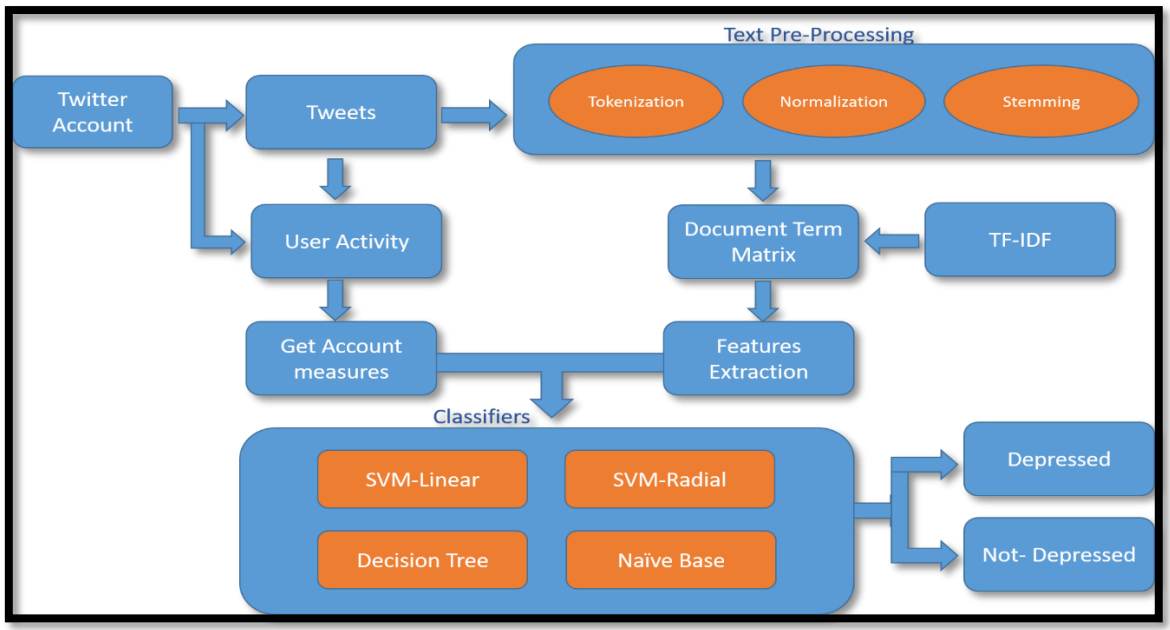

*Figure 3: DDACF classification model*

### 3.1 Feature Engineering

Feature engineering is referred to in machine learning as "the process of using domain knowledge of the data to create features that can be used by machine learning algorithms to find patterns" [17]. Features are generated to extract the information understood by a machine learning algorithm and might be useful for prediction [17]. Types and number of features has significant effect on the efficiency of machine learning algorithms.

Twitter platform has massive amount of information about the user, various features can be extracted from the activity histories and tweets of Twitter users. Features are extracted from the text after text preprocessing, when the text is in the desired format, as shown in Figure (4). These features are computed for both the training and test sets. Table 1 lists the features and their possible values used for the classification model, Where T (true) and F (false) for possible values indicate the use of this feature or not. For example, when the possible value for TF-IDF is T meaning TF-IDF is used for the experiment and if it's F that means word frequency is used instead.

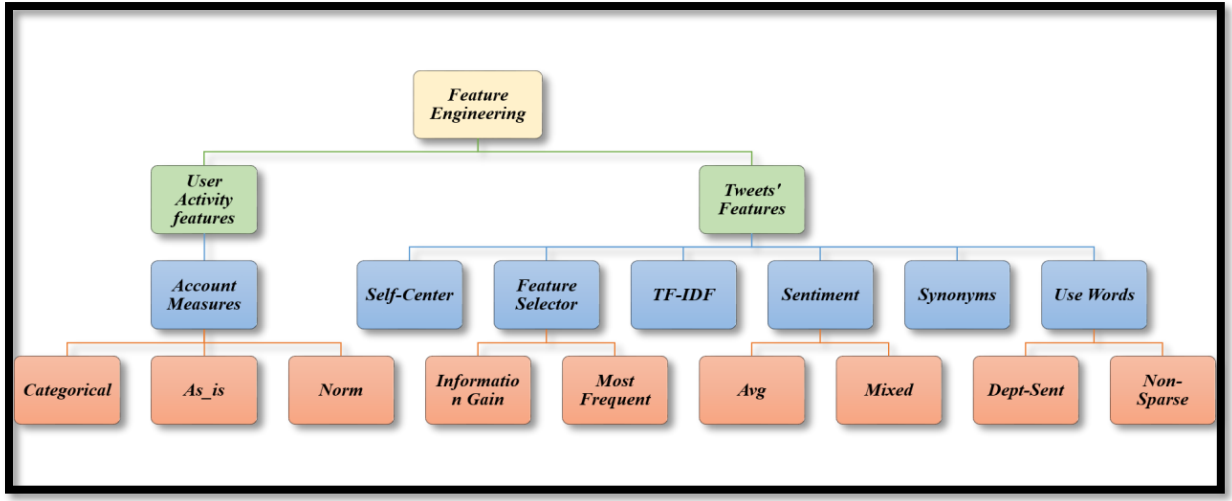

Figure 4: Visualization of features used in the study

Table 1: Description of features and their possible values

| Features | Possible Values | Description |
|---|---|---|
| **Self-Center** | T | Use first-person pronouns |
| | F | Remove all stop words with the first-person pronouns |
| **TF-IDF** | T | Determine the relative frequency of words in a specific document compared to the inverse proportion of that word over the entire document corpus |
| | F | Use word frequency |
| **Feature Selector** | Information Gain (IG) | - Measures the number of bits of information obtained for category prediction by determining the presence or absence of a term in a document.<br>- Words are selected according to the higher IG |
| | Most Frequent | Selects the most frequent words according to the words' higher frequency |
| **Sentiment** | Avg | For each user, sentiment is calculated for each tweet using sentence sentiment, and then, average of all tweets is calculated for the user |
| | Mixed | Selects a higher sentiment for sentences that are negative or positive but have hidden negative indication |
| **Use Words** | Dept-Sent | Sentiment words – positive and negative words – extracted from depressed user's tweets |
| | Non-Sparse | Words having more than 95% zeros are removed (sparse words are removed) |
| **Account Measures** | As-is | User activities taken as it is (No. of posts, Avg of posts a day, Time of posts, No. of replies, No. of mentions, etc.) |
| | Norm | Activities are normalized, and average is calculated according to number of user posts. |
| | Categorical | Activities are categorized according to 4 quartiles (low, below |

|  |  | average, average, and high). |
|---|---|---|
| **Synonyms** | T | Words in the matrix are grouped and frequency is summed according to their synonyms |
|  | F | Words are used as it is, without reducing it according to synonyms |

*Self-Center*
Previous studies have shown that first-person pronouns are useful predictors of depression. De Choudhury and Jamil [8, 17] indicated that the use of singular pronouns in comparison to second- and third-person pronouns is also an indicator of depression. Wang et al. [30] mentioned that depressed users lean more toward the use of first-person singular pronouns but less emoticons. Thus, we skip removing the first-person pronouns with other stop words in the normalization step in the proposed classification model to increase the efficiency of the classification algorithm.

*TF-IDF*
TF-IDF is also added to the classification model, and is used by Ramos, Resnik et al., and De Choudhury et al. [8, 31, 32] to rank words used by users with mental illness [17]. To reduce computation time, sparse vectors are removed from the matrix. Words having more than 95% zeros indicate low use in the user account, and as a result, are removed.

*Feature Selector*
For selecting features there is two possible values, either Most-frequent which selects the most frequent words according to the words' higher frequency or Information gain. Inspired by Prieto et al. [33], information gain (IG) is added as a feature selector for the model. Prieto et al. [33] used IG to reduce features that improve the classification of depressed users by 18% in AUC and 7% in f-measure, and reduced the time needed to generate the model. IG is used in machine learning as a term for goodness criterion. It measures the number of information bits obtained for category prediction by determining the presence or absence of a term in a document. Let $X_i^m = 1$ denote the set of categories in the target space. Then, IG of term *t* is defined as [34]

$$G(y) = -\sum_{i=1}^{m} P_r(x_i) log P_r(x_i) + P_r(y) \sum_{i=1}^{m} P_r(x_i|y) log P_r(x_i|y) + P_r(\bar{y}) \sum_{i=1}^{m} P_r(x_i|\bar{y}) log P_r(x_i|\bar{y})$$

*Sentiment*
Sentence sentiment is used for each tweet in the user's account, then the average of all tweets' sentiment is calculated and this is the Avg feature. Mixed feature calculates sentence sentiment for sentences that are either negative or positive but have hidden negative indication.

*Use Words*
This feature has two possible values, either non-sparse meaning non-sparse words are used and sparse words having more than 95% zeros are removed, or Dept_Sent. Dept_Sent is a feature, inspired by De Choudhury et al. [8], concentrates on depressed users' sentiment words. From tweets crawled for this study, sentiment words, positive and negative, are extracted from depressed users' tweets and put into files and all other words are removed for all users. The exploited feature in this study, Dept_Sent, is distinguished by the fact that it does not use static lexicon words for representing depression. Dept_Sent generalizes the depression lexicon and can be extended easily.

*Account Measures*
Tsugawa et al. [13] showed that features obtained from user activities can be used to predict user depression with 69% accuracy. In addition, De Choudhury [8] used features obtained from the records of individual user activities on Twitter to identify depressed users. Tsugawa et al. and Del Vicario et al. [13, 35] indicated that the more a user is active, the higher is his/her tendency to express negative emotion when commenting, which will help indicate whether the user is depressed.

As a result, aggregated features are used in this paper to help detect depressed users on Twitter. Activities extracted from each user account such as retweets, mentions, ...etc. used in this study are shown in Figure (5).

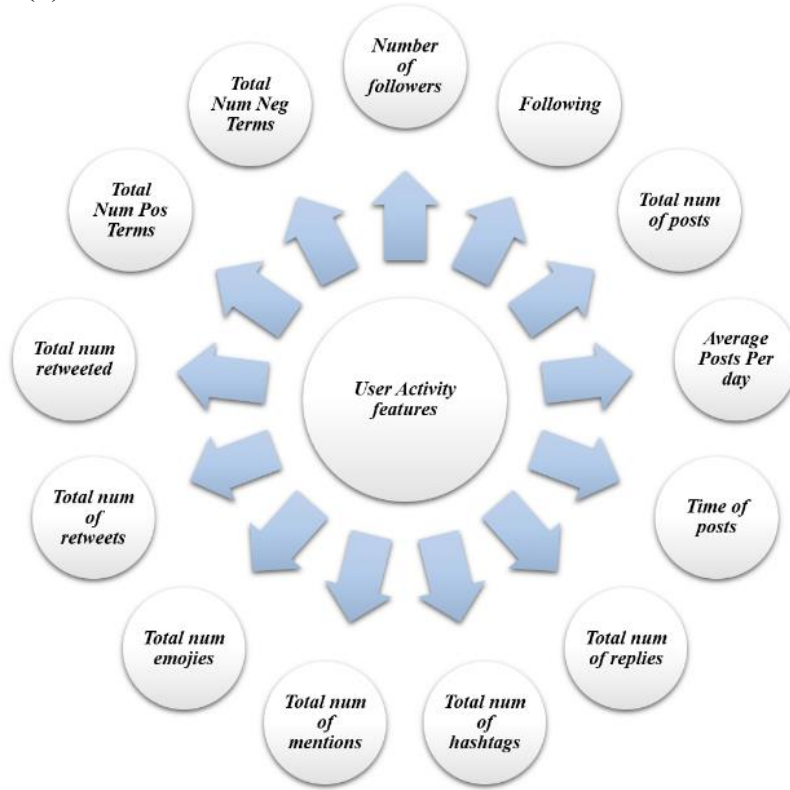

*Figure 5: User activity features extracted from user account*

Three different possible values for this feature (As_is, Norm, Categorical). As_is uses user's activities as it is while Norm uses the activities after calculating the average according to the number of user's posts. Categorical is a new feature that has been introduced uniquely in this study. It relies on categorizing activities of each user into four types (low, below average, average, and high), whose delimiters are defined using percentile values from quartile distribution (Q1, Q2, and Q3).

*Synonyms*

This feature reduces the number of words in the matrix by finding similar words and adding frequencies of synonyms. Tsugawa et al. [13] used the bag-of-words approach to reduce the number of words and found that it helped increase the accuracy. Using WordNet [36], frequencies of synonyms are added; for example, if the depressed frequency is "10," sad is "5," and upset is "3," the frequencies will be added, which will amount to "18," and put under one word. This will make the word stronger for detecting depression and reduce the number of words in the corpus, thus decreasing the computation time.

4.  **Experiment and Results**

### 4.1 Data Collection

We use a dataset for Twitter users who suffer from depression. The self-reports are collected by searching Twitter using a regular expression ("..diagnosed with depression.."). Candidate users are filtered manually, and then, all their recent tweets are continuously crawled using the Twitter Search API. To ensure that users are disclosing their own depression and not talking about a friend or a family member, a human annotator reviews these tweets. For each user, up to 3000 of their most recent public tweets are included in the dataset, and each user is isolated from the others. Note that this 3000-tweet limit is derived from Twitter's archival polices [40]. Non-depressed users are collected randomly and checked manually to ensure they never post any tweet containing the character string "depress." In an effort to minimize noisy and unreliable data, users with fewer than five Twitter posts are excluded.

### 4.2 Evaluation Measures

Critique and cross validation of the feasibility of this automated prediction will be conducted through standard accuracy (Acc), precision (P), recall (R), and F1 scores, as well as confusion matrix (CM), and recipient operating classification curves (ROCs), which are defined as follows:

*Accuracy*: the simplest and mostly used measure to evaluate a classifier. It is defined as the degree of right predictions of a model (or contrarily, the percentage of misclassification errors) [37].

$$Acc = \frac{true\ positive + true\ negative}{true\ positive + true\ negative + false\ positive + false\ negative}$$

*Precision:* is defined as the fraction of correctly classified positives to the total predicted positives [37]. Under our condition, it aims to find how many of the users identified as depressed are actually depressed [10].

$$P = \frac{true\ positives}{true\ postives + false\ postives}$$

*Recall:* is defined as the fraction of correctly classified positives to total positives [37]. Within our situation, it aims to determine that of all depressed users, how many are properly detected [10].

$$R = \frac{true\ positives}{true\ positives + false\ negatives}$$

The trade-off between recall (false negatives) and precision (false positive) is compromised by considering the F1-measure:

*F1 Score (F-measure):* is the harmonic mean of precision and recall; it weighs each metric evenly, and therefore, is commonly utilized as a classification evaluation metric [10, 30].

$$F1 = \frac{2 * P * R}{P + R}$$

Hence, it is important to achieve both high recall and high precision.

*CM:* is a form of contingency table that presents the differences between the true and predicted classes for a set of labeled instances [38]. It has four categories: true positives (TP), which refers to positives that are identified correctly; false positives (FP), which are positives identified incorrectly and supposed to be negatives; true negatives (TN), which refer to negatives that are correctly labeled as negative, and false negatives (FN), correspond to positives that are incorrectly labeled as negative [39].

CM can be used to generate a point in the ROC space using metrics that are defined as [37]

$$CM = \begin{bmatrix} TN & FP \\ FN & TP \end{bmatrix}$$

*ROC:* is recommended for evaluating binary decision problems. ROC illustrates how the number of correctly classified positive instances varies with the number of incorrectly classified negative instances [39]. The ROC assessment technique uses the TP and FP rates, which are defined as [17]

$$\text{TP rate} = \frac{\text{True Positive}}{\text{actual positives}}$$

$$\text{FP rate} = \frac{\text{False Positive}}{\text{actual negatives}}$$

The ROC graph plots the TP rate over the FP rate. The performance of a single classifier on a given distribution is represented by a point in ROC space. The area under the ROC curve (AUC) is frequently used as an evaluation criterion to evaluate different classifiers' performances [17].

### 4.3 Results

The experiments are conducted on all possible combinations of feature values indicated in table (1), using various classification algorithms (SVM with different kernels, DT, and NB). The expected labels for any training/testing sample are depressed/not depressed.

From previous studies, "first-person pronouns" and "TF-IDF" have been proven to be discriminant for depression identification. This finding has been proven during experiments run via measuring the correlation between features and class labels. For obtaining optimal classification results, various feature combinations are exploited. First, account measures (as-is and normalization) with sentiment features (mixed, average, and none of them) are used for training and testing. Figure 4 summarizes the F1 measures for different algorithms.

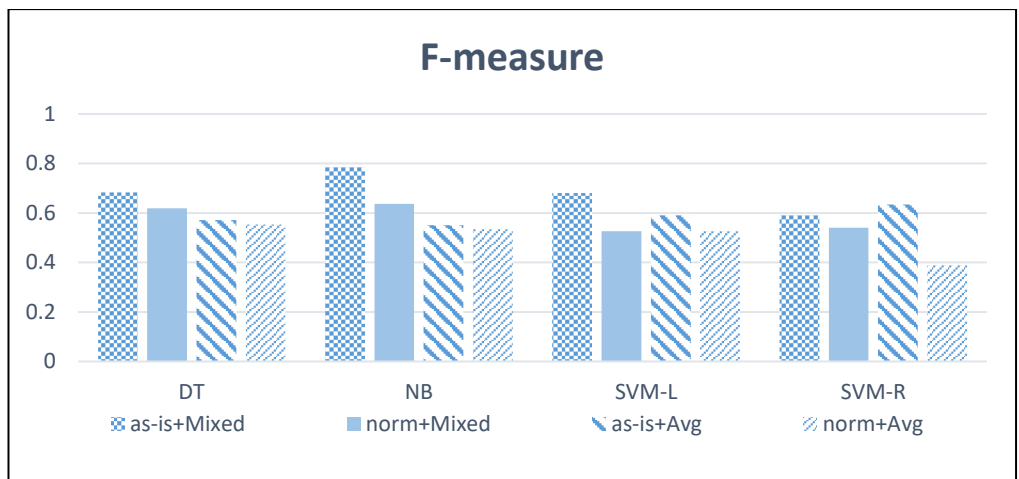

*Figure 6: F1 measures for different algorithms*

Figure 6 shows that using a mixed sentiment with the account measures as they are (as-is) grants a higher F1 measure for all exploited classification algorithms. The F measure is 0.784 for NB and 0.68 for both DT and SVM-L.

In addition, training and testing are conducted to reach out the optimal feature selector (InfoGain and MostFreq). The experimental results show that InfoGain provides higher f-measure results for all classification models as shown in Table 1.

*Table 2: Evaluation measures for different feature selectors.*

| | InfoGain | | | | MostFreq | | | |
|---|---|---|---|---|---|---|---|---|
| | Precision | Recall | f-measure | AUC | Precision | Recall | f-measure | AUC |
| DT | 0.5625 | 0.473684 | 0.514286 | 0.570175 | 0.636364 | 0.388889 | 0.482759 | 0.603535 |
| NB | 0.857143 | 0.352941 | 0.5 | 0.654731 | 1 | 0.3 | 0.461538 | 0.65 |
| SVM-L | 0.619048 | 0.684211 | 0.65 | 0.651629 | 0.6 | 0.521739 | 0.55814 | 0.525575 |
| SVM-K | 0.608696 | 0.7 | 0.651163 | 0.625 | 0.619048 | 0.684211 | 0.65 | 0.651629 |

Another experiment considers the "Dept-Sent" feature for the obtained data, and the results show that the accuracy and f-measure increase for all classifiers, except for DT. The F measure with the DT classifier decreases due to a decrease in precision; meanwhile, the recall increases from 0.56 to 0.69 (less false negatives). In addition, the F measure increases from 0.57 to 0.65 with SVM-L, and shows a slight up-lift with NB and SVM-R, as indicated in Table 3.

*Table 3: Evaluation measures for considering "Dept-Sent" feature.*

| Classifiers | Feature | Accuracy | Precision | Recall | f-measure | AUC |
|---|---|---|---|---|---|---|
| DT | Without Dept-Sent | 75 | 0.764706 | 0.565217 | 0.65 | 0.664962 |
| | With Dept-Sent | 72.5 | 0.473684 | 0.692308 | 0.5625 | 0.660969 |
| NB | Without Dept-Sent | 65 | 0.769231 | 0.454545 | 0.571429 | 0.643939 |
| | With Dept-Sent | 67.5 | 0.571429 | 0.6 | 0.585366 | 0.575 |
| SVM-linear | Without Dept-Sent | 75 | 0.6875 | 0.5 | 0.578947 | 0.611111 |
| | With Dept-Sent | 72.5 | **0.681818** | **0.625** | **0.652174** | **0.59375** |
| SVM-radial | Without Dept-Sent | 75 | 0.631579 | 0.571429 | 0.6 | 0.601504 |
| | With Dept-Sent | 75 | 0.733333 | 0.55 | 0.628571 | 0.675 |
| Without Dept-Sent features are: Self-Center+InfoGain+TF-IDF+Mixed+as_is+non-sparse | | | | | | |
| With Dept-Sent features are: Self-Center+InfoGain+TF-IDF+Mixed+as_is+Dept_Sent | | | | | | |

Table 3 shows the trade-off between recall and precision for each model. SVM shows lower variance between the two measurements.

The next step in the experiment is considering the "Categoral" feature, which is a newly derived feature that has not been considered previously. Exploiting this feature shows a noticeable increase in the F measure with all classification models, except for SVM-L, which shows no change but a slight increase in the recall, as shown in Table 4. These results support those of De Choudhury [8], as he concluded that social media activity provides useful signals that can be utilized to classify whether an individual is suffering or will suffer from depression.

*Table 4: Evaluation measures for considering "Categoral" feature.*

| Classifiers | Features | Accuracy | Precision | Recall | f-measure | AUC |
|---|---|---|---|---|---|---|
| DT | With Dept-Sent | 72.5 | 0.473684 | 0.692308 | 0.5625 | 0.660969 |
| | Dept-Sent and Categ | 70 | 0.631579 | 0.631579 | 0.631579 | 0.649123 |
| NB | With Dept-Sent | 67.5 | 0.571429 | 0.6 | 0.585366 | 0.575 |
| | Dept-Sent and Categ | 75 | 0.684211 | 0.722222 | 0.702703 | 0.724747 |
| SVM-L | With Dept-Sent | 72.5 | 0.681818 | 0.625 | 0.652174 | 0.59375 |
| | Dept-Sent and Categ | 72.5 | 0.666667 | 0.636364 | 0.651163 | 0.623737 |
| SVM-R | With Dept-Sent | 75 | 0.733333 | 0.55 | 0.628571 | 0.675 |
| | Dept-Sent and Categ | 77.5 | **0.736842** | **0.777778** | **0.756757** | **0.775253** |
| With Dept-Sent features are: Self-Center+InfoGain+TF-IDF+Mixed+as_is+Dept_Sent | | | | | | |
| Dept-Sent and Categ features are: Self-Center+InfoGain+TF-IDF+Mixed+Categ+Dept_Sent | | | | | | |

The results support the same conclusion: although the dataset becomes richer, the detection algorithm becomes more stable and the trade-off between precision and recall becomes narrower. Moreover, SVM outperforms other detection algorithms with better accuracy.

Besides the other features, the synonyms feature is also applied to the training and testing sets and the classification models show an increase in the F measure, except for SVM-R. Moreover, DT shows a slight decrease but large increase in accuracy. SVM-L and NB show an increase that reaches 82% accuracy with SVM-L and 80% with NB. The F measure is also increased, reaching 0.79 in SVM-L, as listed in Table 5.

*Table 5: Evaluation measures for considering "Synonyms" feature.*

| Classifier | Features | Accuracy | Precision | Recall | f-measure | AUC |
|---|---|---|---|---|---|---|
| DT | + Categ | 70 | 0.631579 | 0.631579 | 0.631579 | 0.649123 |
|  | Categ + Synom | 77.5 | 0.65 | 0.590909 | 0.619048 | 0.60101 |
| NB | + Categ | 75 | 0.684211 | 0.722222 | 0.702703 | 0.724747 |
|  | Categ + Synom | 80 | 0.653846 | 0.809524 | 0.723404 | 0.66792 |
| SVM-L | + Categ | 72.5 | 0.666667 | 0.636364 | 0.651163 | 0.623737 |
|  | Categ + Synom | 82.5 | **0.73913** | **0.85** | **0.790698** | **0.775** |
| SVM-R | + Categ | 77.5 | 0.736842 | 0.777778 | 0.756757 | 0.775253 |
|  | Categ + Synom | 77.5 | 0.705882 | 0.631579 | 0.666667 | 0.696742 |
| +Categ features are: Self-Center+InfoGain+TF-IDF+Mixed+Categ+Dept_Sent | | | | | | |
| Categ + Synom features are: Self-Center+InfoGain+TF-IDF+Mixed+Categ+Dept_Sent+Synom | | | | | | |

After conducting many experiments with variations in the exploited features, the results emphasize that enriching the model with discriminant features yields better results. Moreover, the SVM-linear classifier shows best results and invariant behavior, despite its extensive performance complexity.

Figure 7 shows the overall impact of including "Synonyms" in the feature set (SVM-linear), which increases the frequency of words used by depressed people and gives the words strength when fed to the classification model. It also reduces the number of words in the corpus, which decreases the computation time.

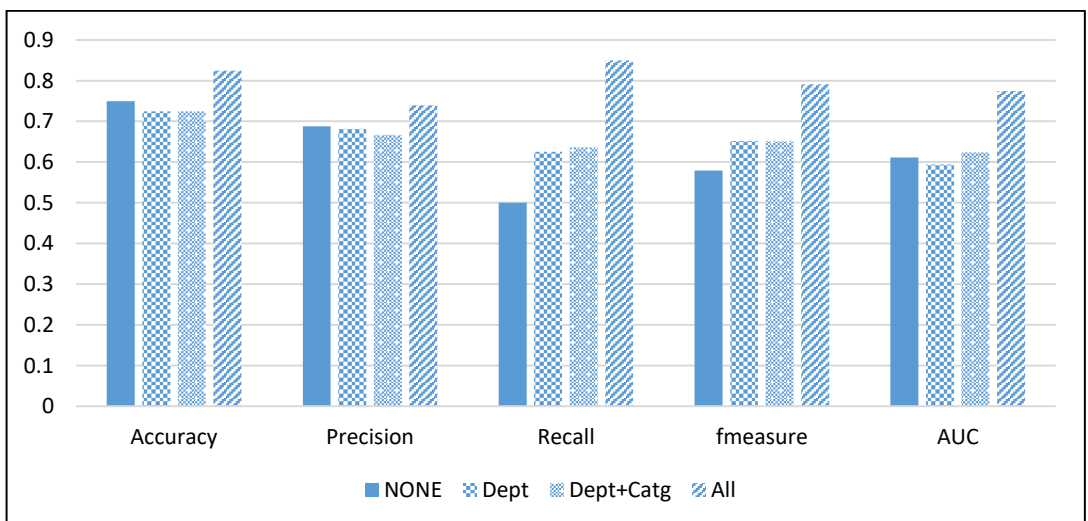

*Figure 7: Results of applying all features to SVM-L*

The main contribution of this study lies in exploiting the concept of merging both tweets' text besides the account historical activities. Both types of features enable the ML models to monitor the change in the mental and psychological states of the user. Table 6 lists a comparative analysis between the proposed model and the previously proposed ones.

*Table 6: Comparative analysis between the proposed model and previously proposed models.*

|  | (De Choudhary 2013)[8] | (Nadeem 2016)[10] | (Reece 2016)[9] | Proposed work |
|---|---|---|---|---|
| **Dataset** | 2M tweets -476 users (questionnaire) | 2.5M tweets (326 depressed) (self-diagnosed) | 105 depressed users out of 204 (CESD scores) | 67 depressed users out of 111 (self-diagnosed) and more than 300,000 tweets |
| **ML models exploited** | SVM | NB | 1200-Random Forest | Linear SVM |
| **Used Features** | - LIWC<br>- Sentiment analysis<br>- Social engagement<br>- Language<br>- Social network | - BOW<br>- Sentiment analysis | - LIWC<br>- Sentiment analysis<br>- Time series<br>- LabMT | -LIWC<br>-Sentiment analysis<br>-Social activity<br>-Synonyms |
| **Obtained Accuracy** | Accuracy: 70% | Accuracy: 81%<br>Precision: 0.86 | Precision: 0.866 | Accuracy:82.5%<br>Recall : 0.85 |

From the previous analysis, we conclude that the proposed model outperforms the previously proposed ones in terms of accuracy, due to the diversity and richness of its feature set. Using features that other studies proved to benefit the classification algorithms and the three new features introduced in this study, classification results outperformed other studies proofing the importance of both the user's activities and tweets to reach good indication of the user's mental health situation.

5. **Conclusion**

This paper defines a binary classification problem as identifying whether a person is depressed, based on his tweets and Twitter profile activity. Different machine learning algorithms are exploited and different feature datasets are explored. Many preprocessing steps are performed, including data preparation and aligning, data labeling, and feature extraction and selection. The SVM model has achieved optimal accuracy metric combinations; it converts an extremely non-linear classification problem into a linearly separable problem. Although the DT model is comprehensive and follows understandable steps, it can fail if exposed to brand-new data. This study can be considered as a step toward building a complete social media-based platform for analyzing and predicting mental and psychological issues and recommending solutions for these users. The main contribution of this study lies in exploiting a rich, diverse, and discriminating feature set that contains both tweet text and behavioral trends of different users. This study can be extended in the future by considering more ML models that are highly unlikely to over-fit the used data and find a more dependable way to measure the features' impact.

**Hatoon S AlSagri** received the Master's degree in information systems from the Department of Information Systems, College of Computer and Information Sciences, King Saud University, where she is currently pursuing the Ph.D. degree in information systems. She is currently a Lecturer with the Department of Information Systems, College of Computer and Information Sciences, Al-Imam Mohammad bin Saud Islamic University. During her graduate studies, she has had the opportunity to participate in various conferences and has published various journal articles. Her main research interests lie in the field of data mining, information diffusion, and social analysis.

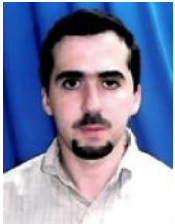
**Mourad Ykhlef** received the B.Eng. degree in computer science from Constantine University, Algeria, the M.Sc. degree was in artificial intelligence from University Paris 13, France, and the Ph.D. degree in computer science from University Bordeaux 1, France. He is currently a Professor with the Department of Information Systems, College of Computer and Information Sciences, King Saud University, Saudi Arabia. His main research interests include data mining, data warehouse, XML and bio-inspired computing.